\documentclass[prl,longbibliography,amssymb,twocolumn,superscriptaddress]{revtex4-1}
\usepackage{amsmath}
\usepackage{amssymb}
\usepackage{amsthm}
\usepackage{amsfonts}
\usepackage{gensymb} 
\usepackage{listings}
\usepackage{enumerate}
\usepackage{latexsym}
\usepackage{psfrag}
\usepackage{bm}
\usepackage[all]{xy}
\usepackage{graphicx}
\usepackage{subfigure}
\usepackage[pdftex,colorlinks]{hyperref}
\usepackage{color}
\usepackage{mathtools}
\usepackage{times}

\begin{document}
\title{Core-level x-ray photoemission and Raman spectroscopy studies on
  electronic structures in Mott-Hubbard type nickelate oxide NdNiO$_2$}
\author{Ying Fu}
\thanks{These two authors contributed equally.}
\affiliation{Shenzhen Institute for Quantum Science and Engineering, and
  Department of Physics, Southern University of Science and Technology, Shenzhen
  518055, China}
\affiliation{Institute of Applied Physics and Materials Engineering, University
  of Macau, Macau 999078, China}
\author{Le Wang}
\thanks{These two authors contributed equally.}
\affiliation{Shenzhen Institute for Quantum Science and Engineering, and Department of Physics, Southern University of Science and Technology, Shenzhen 518055, China}
\author{Hu Cheng}
\author{Shenghai Pei}
\author{Xuefeng Zhou}
\author{Jian Chen}
\author{Shaoheng Wang}
\affiliation{Shenzhen Institute for Quantum Science and Engineering, and
  Department of Physics, Southern University of Science and Technology, Shenzhen
  518055, China}

\author{Ran Zhao}
\affiliation{School of Advanced Materials, Shenzhen Graduate School Peking University, Shenzhen 518055, China}
\author{Wenrui Jiang}
\author{Cai Liu}

\author{Mingyuan Huang}
\affiliation{Shenzhen Institute for Quantum Science and Engineering, and Department of Physics, Southern University of Science and Technology, Shenzhen 518055, China}

\author{XinWei Wang}
\affiliation{School of Advanced Materials, Shenzhen Graduate School Peking University, Shenzhen 518055, China}
\author{Yusheng Zhao}
\author{Dapeng Yu}
\author{Fei Ye}
\author{Shanmin Wang}
\email{wangsm@sustech.edu.cn}

\author{Jia-Wei Mei}
\email{meijw@sustech.edu.cn}
\affiliation{Shenzhen Institute for Quantum Science and Engineering, and Department of Physics, Southern University of Science and Technology, Shenzhen 518055, China}

\date{\today}

\begin{abstract}
We perform core-level X-ray photoemission spectroscopy (XPS) and electronic Raman scattering studies of electronic structures and spin
fluctuations in the bulk samples of  the nickelate oxide
NdNiO$_2$. According to Nd $3d$ and O $1s$ XPS spectra, we conclude that
NdNiO$_2$ has a large transfer energy. From the analysis of the main line of the
Ni $2p_{3/2}$ XPS, we confirm the NiO$_2$ planes in NdNiO$_2$ are of Mott-Hubbard
type in the Zaanen-Sawatzky-Allen scheme. The two-magnon peak in the Raman scattering
provides direct evidence for the strong spin-fluctuation in NdNiO$_2$. The peak
position determines the antiferromagnetic exchange $J=25$~meV. Our experimental
results agree well with our previous theoretical results. 
\end{abstract}

\maketitle

\emph{Introduction --}
Nickel has the formal valences Ni$^{3+}$, Ni$^{2+}$ and Ni$^{+}$ with electronic configurations of of 3$d^7$
, 3$d^8$ and 3d$^9$, respectively, in the transition-metal oxides. The nickelate
compounds exhibit lots of strongly
correlated electronic properties, such as the metal-insulator transitions, unusual magnetic order, charge
order and even
superconductivity~\cite{Zaanen1985,Medarde1997,Catalan2008,Middey2016,Catalano2018,Crespin1983,Hayward1999,Hayward2003,Anisimov1999,Lee2004,Kawai2009,Kaneko2009,Kawai2010,Ikeda2013,Ikeda2016,Li2019,Hepting2019}. The recent discovery of the superconductivity in the thin film
Nd$_{0.8}$Sr$_{0.2}$NiO$_2$~\cite{Li2019} has the similar crystal 
 and  3$d^9$ electronic structures to the cuprate 
superconductors~\cite{Bednorz1986}, and motivates the experimental~\cite{Hepting2019} and lots of theoretical
works~\cite{Botana2019, Sakakibara2019, Jiang2019, Wu2019,
  Nomura2019, Gao2019, Zhang2019,
  Ryee2019,Zhang2020,Zhang2019a,Jiang2019a,Hu2019,Chang2019,Werner2020,Liu2019,Choi2020,Karp2020}
to explore its electronic properties. The recent  superconductivity dome  spanning the doping concentration 0.125$<x<$0.25
in thin-film Nd$_{1-x}$Sr$_x$NiO$_2$ infers the remarkable similarity to cuprate superconductors~\cite{Li2020}.

Reduced form of perovskite nickelate LaNiO$_3$ and NdNiO$_3$
leads to the infinite layered phase
LaNiO$_2$ and
NdNiO$_2$~\cite{Crespin1983,Hayward1999,Hayward2003,Kawai2009,Kaneko2009,Ikeda2013,Ikeda2016}.
Several experimental studies have been carried out on the parent superconducting
nickelate oxides.
Magnetization measurements and powder neutron diffraction were performed on
LaNiO$_2$ and NdNiO$_2$ to study the magnetic
properties~\cite{Hayward1999,Hayward2003}. 
Electric transport
 measurements have been performed on the thin film samples LaNiO$_2$ and
 NdNiO$_2$, and the resistivity exhibits the insulator or semiconductor behavior
 depending on the samples~\cite{Kawai2009,Kaneko2009,Ikeda2013,Ikeda2016,Li2019}. The Ni-K edge X-ray
 absorption spectroscopy (XAS) was
 implemented to characterize the LaNiO$_2$ samples~\cite{Crespin1983,Kawai2009}. Recently, with the help of density functional theory within LDA+U
 scheme, Hepting \textit{et. al.}~\cite{Hepting2019} has used XAS and X-ray
 emission spectroscopy (XES) to unveil a large charge-transfer energy, and
 categorized LaNiO$_2$ and NdNiO$_2$ to the Mott-Hubbard
 type according to the Zaanen-Sawatzky-Allen (ZSA) scheme~\cite{Zaanen1985}.
 The Mott-Hubbard scenario is still  under debate, and the
 charge-transfer type is also proposed  for NdNiO$_2$~\cite{Karp2020}. %

In this paper, the main purpose to study electronic structures in NdNiO$_2$ is twofold. Firstly, we perform the core-level
x-ray photoemission spectroscopy (XPS) to examine the existence of the
Zhang-Rice singlet~\cite{Zhang1988} to address the ZSA classification issue for NdNiO$_2$. Ni$^+$ has higher chemical potential and hence a larger transfer energy in
NdNiO$_2$ as observed in XAS and XES~\cite{Hepting2019}. We implement the
XPS measurement to further disprove the large-$\Delta$ charge-transfer
(a charge-transfer insulator with a large transfer energy $\Delta$) scenario for
NdNiO$_2$. The main line of the transition-metal $2p$ XPS spectrum
has the doubly peaked structure in the cuprate parent compound~\cite{Fujimori1987,Shen1987} and NiO~\cite{Veenendaal1993} due to
the non-local screening effects~\cite{Veenendaal1993,Veenendaal2006,Ghiasi2019},  
related to the Zhang-Rice singlet in charge-transfer
insulators~\cite{Zhang1988}. In the XPS measurements, from Nd 3$d$ and O 1$s$
XPS, we find NdNiO$_2$ has a larger transfer energy $\Delta$, comparing to a
small/negative charge-transfer insulator NdNiO$_3$. 
The non-local screening doubly peaked
structure of the main line of Ni 2p$_{3/2}$ is significantly suppressed in
NdNiO$_2$, confirming the Mott-Hubbard scenario for NdNiO$_2$~\cite{Hepting2019}.

Secondly, as the NiO$_2$ planes in NdNiO$_2$ is in the Mott-Hubbard regime in the ZSA
scheme,  we are concerned with the spin dynamics in the system. The strong
antiferromagnetic spin fluctuation is generally accepted as a key ingredient for the
unconventional superconductivity in the cuprate
oxides, due to the proximity to a Mott insulator~\cite{Anderson1987,Lee2006}. Raman spectroscopy probes spin fluctuations in
the transition-metal compounds, even without a
long-range magnetic order~\cite{Devereaux2007}. The exchange interaction $J$ for cuprates is
very large,  and $J$ was determined from the two-magnon peak
to be of order
125~meV~\cite{Sugai1988,Lyons1988,Lyons1989,Knoll1990,Sulewski1990,Sulewski1991,Blumberg1994,Blumberg1996,Ruebhausen1997,Ruebhausen1999,Sugai2003,Devereaux2007}. The absence of long-range magnetic
order in LaNiO$_2$ and NdNiO$_2$ in Refs.~\cite{Hayward1999} may be due to poor sample qualities, or due to the ``self-doping'' effects
of the Nd $5d$ electron pockets~\cite{Lee2004,Hepting2019,Zhang2019}.  We
perform Raman measurements to explore spin
fluctuations, and observe a broad two-magnon peak in NdNiO$_2$. The antiferromagnetic exchange is determined as $J=25$~meV in NdNiO$_2$,
 close to our previous theoretical estimation~\cite{Zhang2019}.

\emph{Experimental setup. --}
We carried out the XPS investigations on Thermo Fisher ESCALAB 250Xi using
monochromated Al K$\alpha$ radiation at room temperature, and the electron flood gun was
turned on to eliminate electric charging effect in our insulating samples. The
binding energy in XPS was calibrated by $1s$ spectra of carbon. The Shirley
background of the XPS spectra has been subtracted in this work. The energy
dispersive X-ray (EDX) and scanning electron microscopy (SEM) studies were performed on Phenom ProX equipped with
Backscatter detector at an operating voltage of 15 kV.
X-ray diffraction (XRD) measurements were conducted on Rigaku Smartlab 9~KW using Cu K$\alpha$
radiation at room temperature. The Raman spectra were measured in the quasi-back-scattering geometry on our home-built system 
using a HORIBA iHR550 spectrometer and the 632.8~nm excitation line of a He-Ne
laser. The power of the laser is about 400~$\mu$W, and we set 1200~grooves/mm
grating and 30~min integral time. The samples were
placed in a He-flow cryostat which evacuated to $2.0\times10^{-6}$~Torr. 

\emph{Single crystal synthesis. --}
\begin{figure}[t]
  \centering
  \includegraphics[width=\columnwidth]{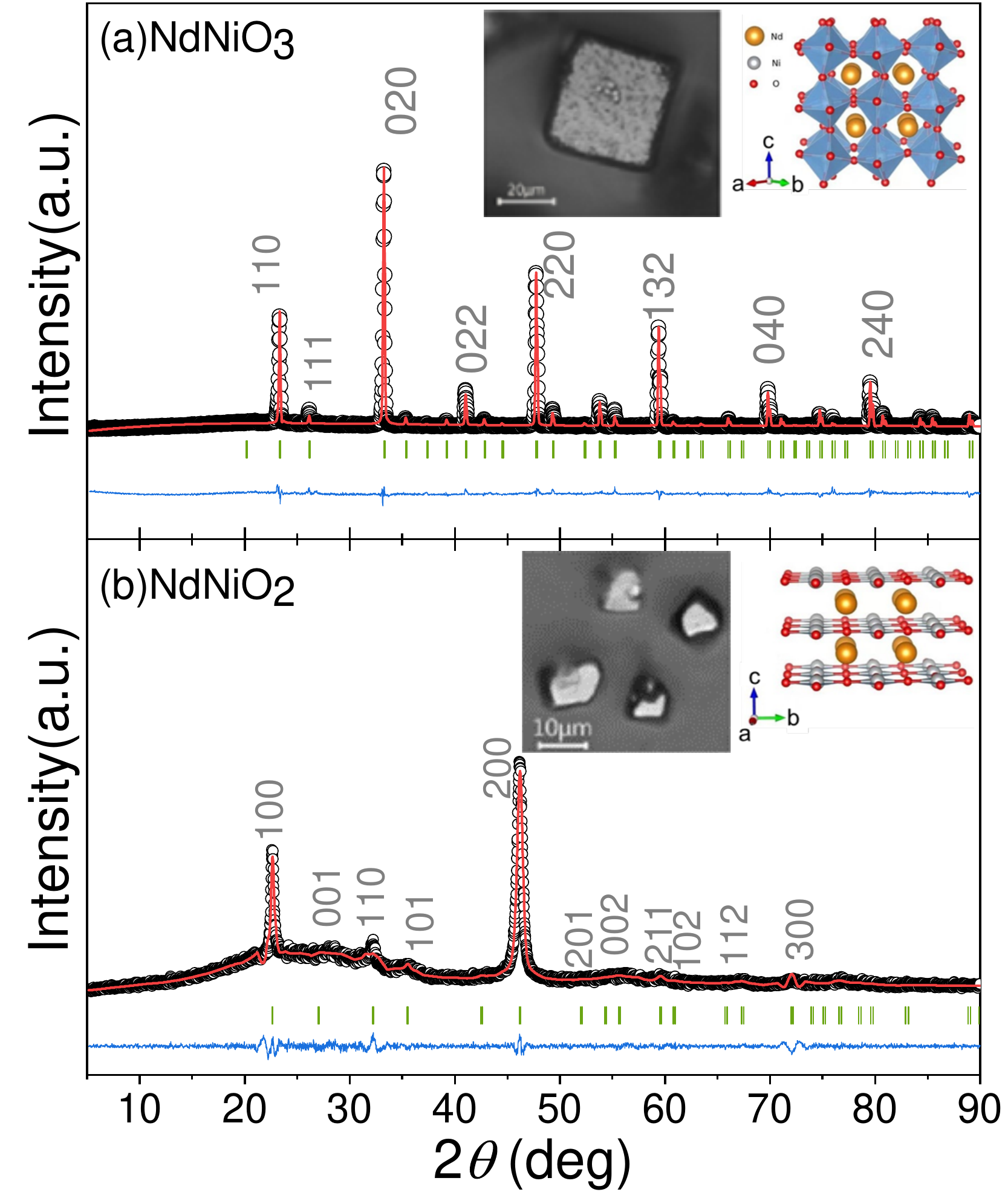}
  \caption{Refined XRD patterns collected at ambient conditions using a Cu
    K$\alpha$ radiation. (a) NdNiO$_3$. (b) NdNiO$_2$. The observed and
    calculated patterns are shown in black and red, respectively. The positions
    of allowed Bragg reflections are denoted by green tick marks. The bottom
    profiles are the difference between the observed and calculated results.
    Insets of each panel are crystal pictures and the polyhedral view of the corresponding crystal structures. }
  \label{fig:figure1}
\end{figure}
The NdNiO$_3$ polycrystals were synthesized by a combination precursor method with
high pressure technique. The precursor was prepared by sintering the sol-gel
with stoichiometric ratio of Ni and Nd~\cite{Vassiliou1989}. After calcining the
precursor powder capsuled in h-BN at 1073 K under high pressure of 2~GPa, the
crystals of NdNiO$_3$ with high quality were successfully obtained, as shown in
Fig.~\ref{fig:figure1}. The reduction of NdNiO$_3$ was performed at low temperature
using CaH$_2$ as the
reducing agent, as reported in refs~\cite{Kawai2009,Li2019}, and finally, the $P4/mmm$ phase of
NdNiO$_2$ was obtained.

We observe the broadening of the peak width in the XRD pattern for NdNiO$_2$, especially in the
peak indexed by $\langle hkl \rangle$ with nonzero value for $l$. The anisotropic broadening of the Bragg
peaks  was also observed in both thin films and
polycrystals~\cite{Hayward1999,Hayward2003,Kawai2009,Kaneko2009,Li2019,Li2019a},
and could be ascribed to the anisotropic crystal size or stacking disorder. 
NdNiO$_2$ has intensive peaks along $\langle h00 \rangle$  direction in XRD pattern as shown in Fig.~\ref{fig:figure1} (b), probably
due to the preferred $a$-axis of crystal grains. The orientation preference was
reported for the thin-film LaNiO$_2$ where the preferred orientation changes
from $c$-axis to $a$-axis for the over-time reduction~\cite{Kaneko2009,Kawai2010}. 

We performed the Rietveld profile refinement for NdNiO$_3$ and NdNiO$_2$ using
the Fullprof suite of programs~\cite{RodriguezCarvajal1993}. The NdNiO$_3$ has
the space group of $Pbnm$ with lattice parameters $a = 5.38500$~\AA, $b =
5.38400$~\AA, and $c =7.61314$~\AA~ ($R_{wp}=8.83$\% and $R_B=13.25$\%),
consistent with the previous report~\cite{Garcia-Munoz2009}.  The refinement
results for NdNiO$_2$ is $a = b = 3.93110$~\AA~ and  $c = 3.30200$~\AA~ ( with
$R_{wp}= 11.6$\% and $R_B=7.35$\% ), which is in agreement with previous results
which used NaH as reduction agent~\cite{Hayward2003}. 
It is worth noting that the metal nickel phase usually appears in NdNiO$_2$ after reduction, as
reported in refs~\cite{Hayward1999,Hayward2003,Kawai2009,Li2019a}. Although the
metal nickel phase is almost invisible in our
PXRD result, we observed the metal nickel impurities embedded in surface-polished
NdNiO$_2$ crystal by SEM, as shown in inset of Fig.~\ref{fig:figure5}.

\begin{figure}[b]
  \centering
  \includegraphics[width=\columnwidth]{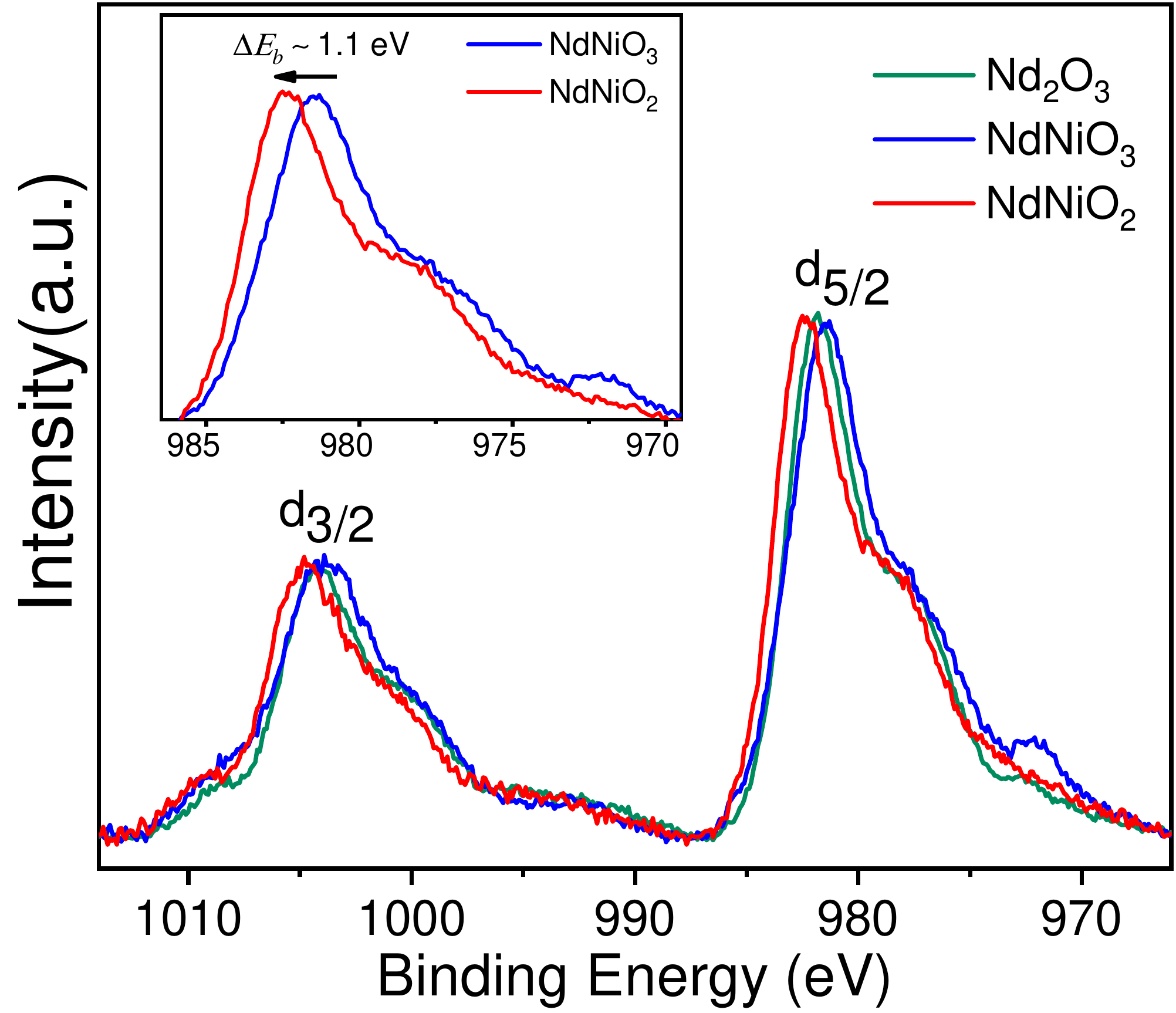}
  \caption{Nd $3d$ core-level XPS spectra of NiO, NdNiO$_3$ and NdNiO$_2$. 
  }
  \label{fig:figure2}
\end{figure}
\emph{Electronic structures from XPS. --}
The core-level x-ray spectroscopy is a powerful tool
to study electronic properties of transition-metal
oxides~\cite{DeGroot2008,VanVeenendaal2015}. It helps to determine the on-site Coulomb interaction and
the charge-transfer
energy, and classify materials as charge-transfer or
Mott-Hubbard insulators in the Zaanen-Sawatzky-Allen scheme~\cite{Laan1981,Gunnarsson1983,Zaanen1985,Zaanen1986,Bocquet1992}.

Figure~\ref{fig:figure2} is the Nd 3$d$ core-level XPS spectra for Nd$_2$O$_3$,
NdNiO$_3$ and NdNiO$_2$. The similar overall pattern indicates the trivalent
Nd$^{3+}$ with the 4$f^3$ ground state in the
three compounds. There is a blue shift of the binding energy around $\Delta
E_b\simeq1.1$~eV for NdNiO$_2$, comparing to NdNiO$_3$ as shown in the inset. 
Thus the formal chemical valence is Ni$^{3+}$ and Ni$^{1+}$ in
NdNiO$_3$ and NdNiO$_2$, respectively. In the Nd$^{3+}$ $3d$ XPS, the spectral
structure is mainly determined by the effect of the covalency hybridization, and
the two peaks of the $3d$ XPS correspond to the bonding and
antibonding states between the $3d^94f^3$ and $3d^94f^4\underline{L}$
configurations (($\underline{L}$ is the
hole in the ligand) in the final states~\cite{DeGroot2008}. 
The ground state of Nd$_2$O$_3$ is in the almost pure 4$f^{3}$
configuration, and hence the main peak at around 981-983~eV in the Nd 3$d_{5/2}$ XPS spectra is the
unscreened $3d^94f^3$ final state. The screened $3d^94f^4\underline{L}$ final state
has a lower binding energy around 977~eV. 





\begin{figure}[t]
  \centering  
  \includegraphics[width=\columnwidth]{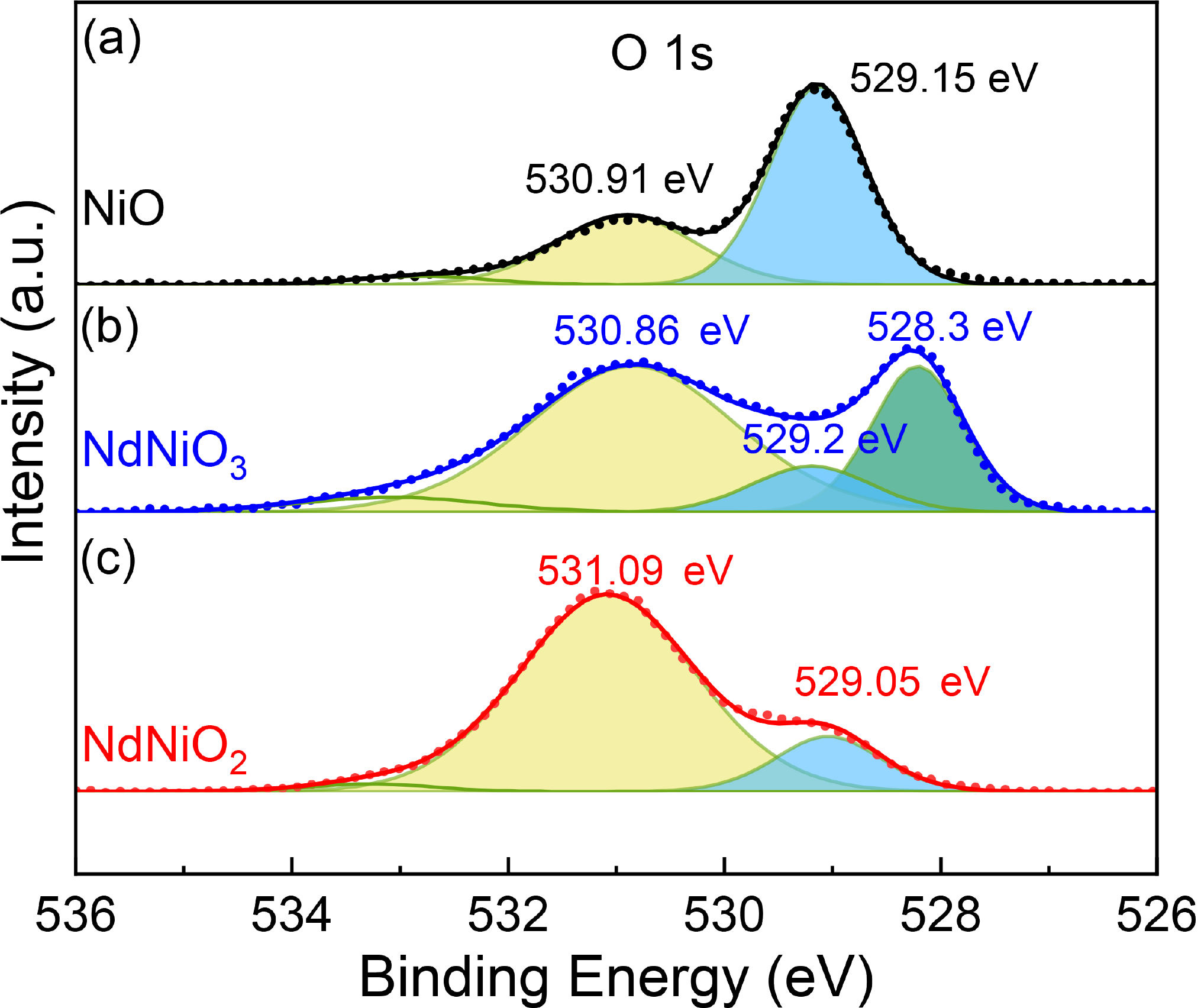}  
  \caption{O 1s core-level XPS spectra of NiO (a), NdNiO$_3$ (b) and NdNiO$_2$
    (c) powder samples.}
  \label{fig:figure3}
\end{figure}

Figure~\ref{fig:figure3} compares the O 1$s$ XPS spectra in NiO, NdNiO$_3$ and
NdNiO$_2$. The O 1$s$ XPS has often used in the experiments for the transition-metal
oxides, however, the nature of the 531~eV O 1$s$ spectral peak in NiO
is unclear~\cite{Sarma1990,Biesinger2009}, and has been proposed to be due to
defective sites within the oxide crystal~\cite{Hagelin-Weaver2004,Payne2009},
adsorbed oxygen~\cite{Benndorf1982}, or hydroxide species~\cite{Carley1983}, and
also attributed to an intrinsic feature~\cite{Weaver1988,Sarma1990}. We can't exclude the contaminant origin for
the 530.9~eV feature, however, we have measure the O 1$s$ XPS spectra on
both different samples and different XPS instruments and obtain the consistent similar
results, which is hardly attributed to an extrinsic feature. 
  In O $1s$ XPS spectra, NiO
and NdNiO$_2$ have the best-screened states at around 529~eV. In NdNiO$_3$,
an extra peak appears at the lowest binding energy 528.3~eV, indicating the
concentration of the oxygen holes in the ground state, due to a negative
charge transfer energy~\cite{Johnston2014,Mizokawa1991,Mizokawa1995,Abbate2002}.


As depicted in the inset of Fig.~\ref{fig:figure2}, NdNiO$_2$ has a higher binding energy than NdNiO$_3$ in the
Nd $3d$ XPS by around $\Delta E_b\simeq1.1$~eV. The Nd$^{3+}$ binding energy blue shift $\Delta E_b$
indicates the increasing of the Fermi energy in NdNiO$_2$. Since Nd$^{3+}$ has a
weaker hybridization than Ni$^+$ with the oxygen, the Fermi energy increasing is mainly
due to the increasing chemical potential of Ni$^+$ in
NdNiO$_2$, assuming the unchanged chemical potential of oxygen. 
Hence NdNiO$_2$ has a larger charge-transfer
energy $\Delta$. Taking account for the covalency of the transition-metal ions and
ligands, the effective charge transfer energy is $\Delta'=\Delta-(W+w)/2$~\cite{Bisogni2016},
where $W$ and $w$ are the bandwidths for the transition-metal 3$d$-band and oxygen
2$p$-band, respectively. NdNiO$_2$ has the nearly two-dimensional transition-metal 3$d$-band and oxygen
2$p$-band which are narrower than the three-dimensional ones in NdNiO$_3$; then
it has has a larger
effective charge transfer energy $\Delta'$. From the comparison between
NdNiO$_2$ and NdNiO$_3$, the former has a large charge-transfer energy than
the latter, and is not a negative/small charge-transfer insulator.
NdNiO$_2$ has the quite similar pattern of O 1$s$ XPS to NiO, but the covalence
peak is significantly suppressed as shown in Fig.~\ref{fig:figure3}.  Since NiO
is a three-dimensional charge-transfer insulator, NdNiO$_2$ could have a larger effective
charge-transfer energy $\Delta'$ than NiO, and is likely to be a
 Mott-Hubbard type as suggested in
 Ref.~\cite{Li2019,Hepting2019,Jiang2019,Zhang2019}, which need further
 confirmation in the Ni $2p$ XPS as demonstrated below.

\begin{figure}[b]
  \centering
  \includegraphics[width=\columnwidth]{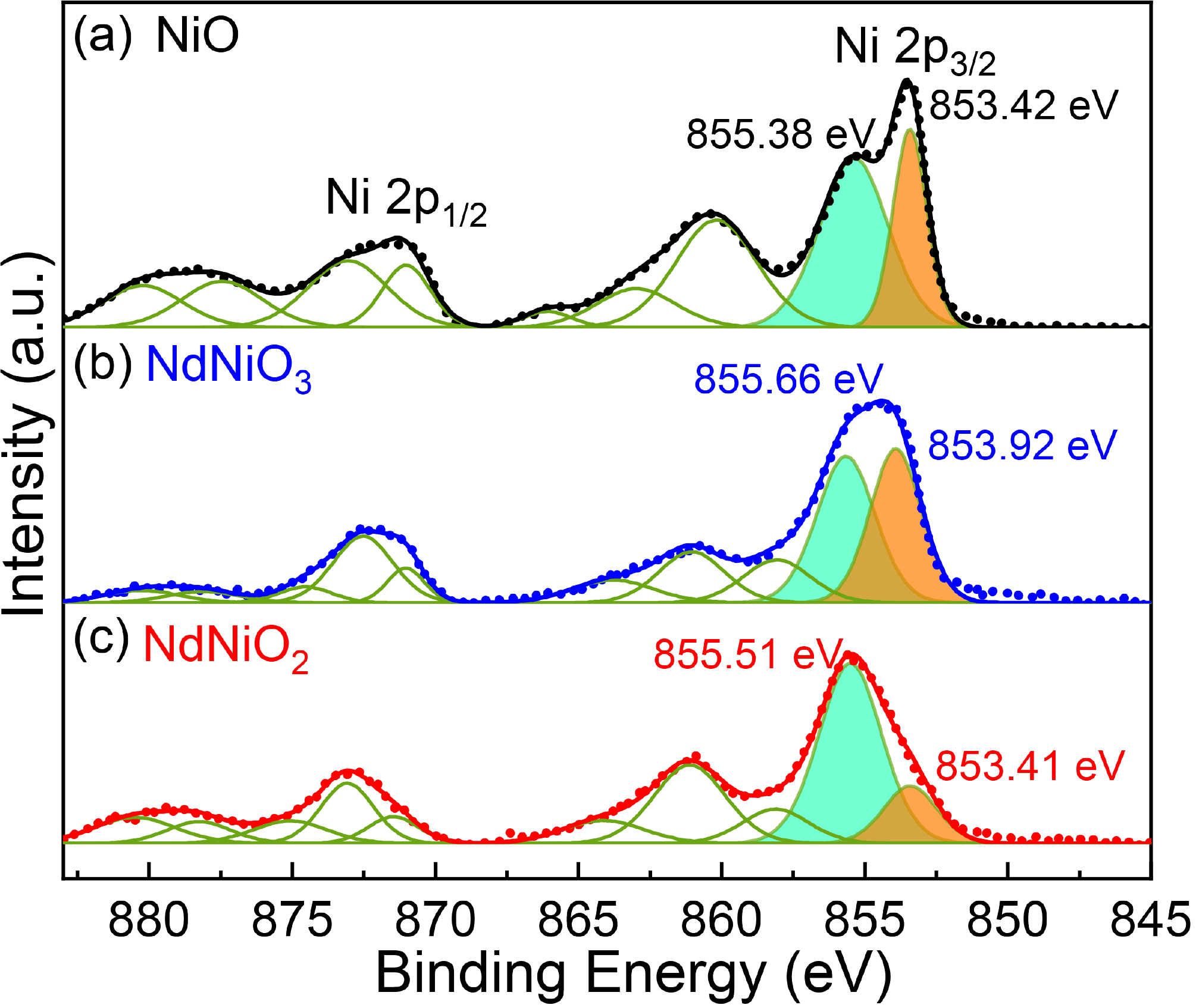}    
  \caption{Ni 2$p_{1/2}$ and 2$p_{3/2}$ core-level XPS spectra of NiO (a), NdNiO$_3$ (b) and NdNiO$_2$ (c) powder samples.}
  \label{fig:figure4}
\end{figure}
Next, we examine the Ni 2$p$ XPS spectra to confirm whether NdNiO$_2$ is a
charge-transfer or Mott-Hubbard type. 
Figure~\ref{fig:figure4} displays Ni 2$p$ XPS spectra of NiO, NdNiO$_3$ and
NdNiO$_2$. The Ni 2p spectrum is composed of 2$p_{3/2}$
and 2$p_{1/2}$ components due to the large spin-orbit coupling in the Ni $2p$
core level. 
The main line  region of the $2p_{3/2}$ spectra around 852-859~eV corresponds to the screened
final states, and can be fitted by two
peaks. 
The removal of the 2$p$ core electron leads to the creation of a strong local 
potential on the nickel site, and an electron is transferred  from the
environment to the Ni site. The different types, i.e., small/negative
charge-transfer, charge-transfer and Mott-Hubbard types according to the ZSA
scheme, of the transition-metal compounds have different non-local screening effects in the main line of the $2p$ transition-metal
XPS spectra~\cite{Veenendaal1993,Veenendaal2006}.


The perovskite NdNiO$_3$ has a negative effective charge-transfer energy, giving
rise to ``self-doped'' holes on the ligands~\cite{Bisogni2016}. It mainly has a local
Ni 3$d^8$ configuration, a predominant O 2$p$ character across the Fermi level
and a consequent ground state of mainly Ni 3$d^8\underline{L}$ in the metallic
state at the room temperature. The two peaks in the main lines correspond to the
$2p^53d^8\underline{L}$ and $2p^53d^9\underline{L}^2$ final
states~\cite{Mizokawa1995}. NdNiO$_2$ has a larger transfer energy $\Delta$ than
NdNiO$_3$ as demonstrated above according to Fig.~\ref{fig:figure2} and
Fig.~\ref{fig:figure3}, and hence displays a different shape of the main line of Ni
$2p_{3/2}$ XPS. Again, we confirm NdNiO$_2$ is not a negative charge-transfer insulator.


In the charge-transfer insulator NiO, the main line have the double-peak
structure. The higher binding energy peak corresponds
to the $2p^53d^{9}\underline{L}$ final state, 
where an electron is transferred
from the neighboring oxygen atoms (local screening) leaving a hole in the oxygen
ligand orbital $L$.  The peak at the lowest energy corresponds to the final state
where an electron is transferred from oxygen in the neighboring NiO$_6$
octahedron, thereby creating a Zhang-Rice singlet in the farway octahedron (non-local
screening)~\cite{Zhang1988,Veenendaal1993,Veenendaal2006}.

The non-local screening peak is a signature for the presence of the Zhang-Rice
singlet in the charge-transfer insulators, as observed in the cuprate oxides and
NiO~\cite{Fujimori1987,Shen1987,Veenendaal1993,Veenendaal2006}. Compared with NiO, the non-local screening peak in NdNiO$_2$ is
significantly suppressed. Hence the NiO$_2$ layer in NdNiO$_2$ is not a large-$\Delta$
charge-transfer type, but a Mott-Hubbard one. The initial ground state and the final
screened state are mainly $3d^9$ and $3d^{10}\underline{L}$, respectively.



\emph{Two-magnon mode in Raman spectroscopy. --}
\begin{figure}[t]
  \centering
  \includegraphics[width=\columnwidth]{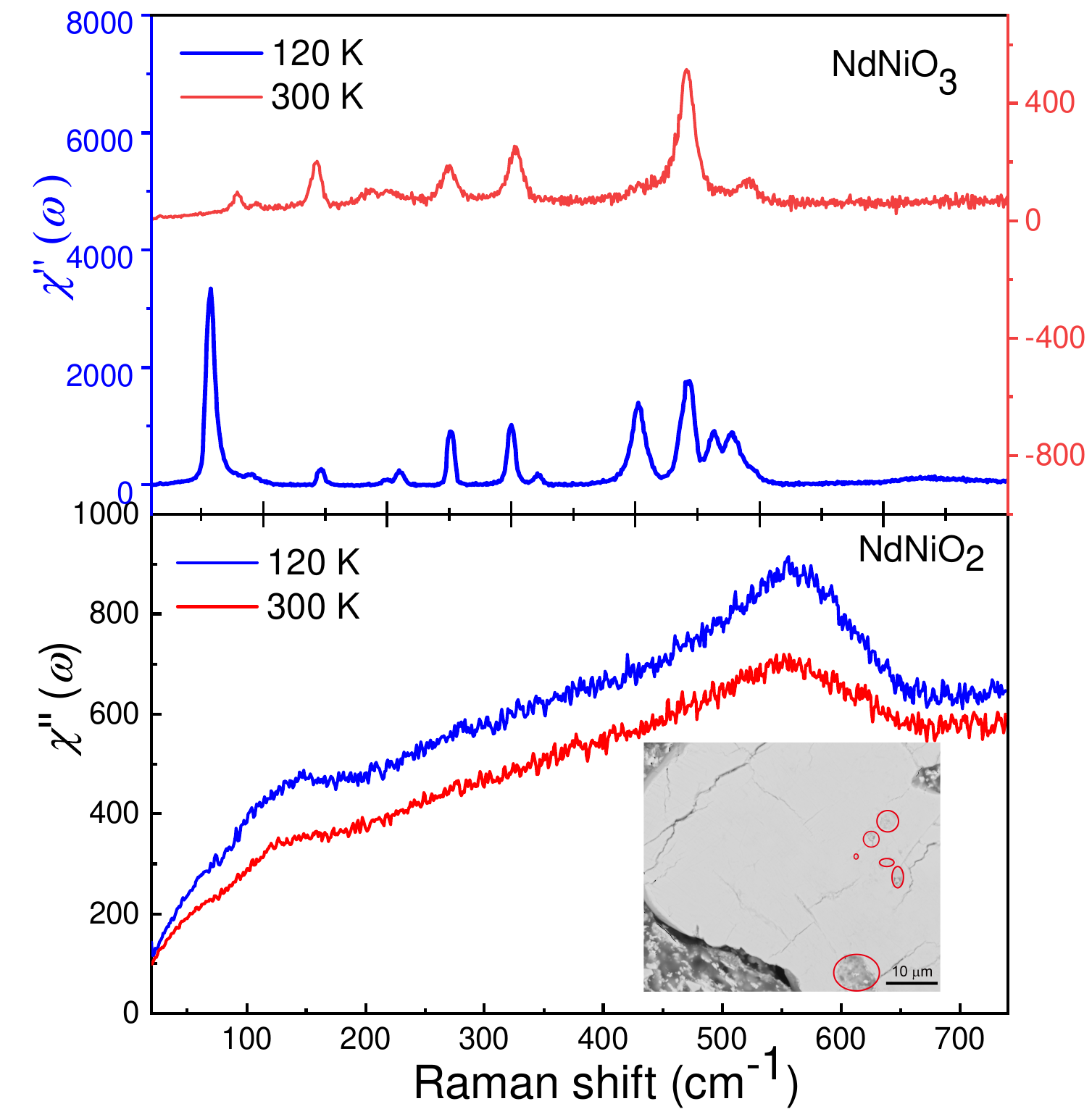}
  \caption{Raman susceptibility in NdNiO$_3$ (upper plane) and NdNiO$_2$ (lower
    plane) at 300~K and 120~K, respectively. The inset is the SEM picture of a
    surface-polished single crystal. The metal nickel has been circled by red
    lines in the inset.
}
  \label{fig:figure5}
\end{figure}
Figure~\ref{fig:figure5} is the Raman spectra of NdNiO$_3$ and NdNiO$_2$ at
300~K and 120~K measured on the sub-millimeter sizable single crystals.  The
Raman susceptibility $\chi''(\omega)$ deduced from the Raman intensity
$I(\omega)$ according to the fluctuation-dissipation theorem,
$I(\omega)=(1+n_B(\omega))\chi''(\omega)$ where $n_B(\omega)$ is the boson
factor.

The Raman spectra for NdNiO$_3$ agrees well with the previous measurements on the thin-film
single crystal~\cite{Zaghrioui2001}, implying the high-quality for our bulk NdNiO$_3$ samples. There are more phonon peaks at low
temperatures due to the structure distortion during the metal-insulator transition around 200~K in
NdNiO$_3$. The dramatically increasing of the phonon intensity is due to the
metal-insulator transition in NdNiO$_3$.

The surface contaminant due to the formation of fluorite NdNiO$_x$H$_y$ 
reported in the previous thin-film NdNiO$_2$~\cite{Onozuka2016,Lee2020} is
nearly invisible in our Nd $3d$ XPS. To remove the possible contaminant, we
polish the NdNiO$_2$ single crystal as shown in the inset of
Fig.~\ref{fig:figure5}. During the synthesis process,  some pure nickel metal are embedded in NdNiO$_2$
crystal due to the perotectic phenomenon and forms the dark gray spots (marked
by red circles) in the
polished surface of NdNiO$_2$.  The
large light gray area in the SEM image is NdNiO$_2$ as identified by EDX. We
perform the Raman scattering measurements on NdNiO$_2$ in the light gray area.

The ideal stoichiometric NdNiO$_2$ has the space group $P4/mmm$ and all the ions
can be taken as the inversion center of the lattice. Therefore, there is no
Raman active phonon mode in the Raman scattering. There is no sharp phonon modes
in the spectra. A broad peak appears at 122~cm$^{-1}$, independent of the
temperature and its origin is unclear. We suspect it
comes from the defect due to the off-stoichiometry of NdNiO$_2$.  

We attribute the mode at around 550~cm$^{-1}$ in the Raman susceptibility spectra of NdNiO$_2$ in Fig.~\ref{fig:figure5} to the two-magnon peak in the two-dimensional Heisenberg antiferromagnet. The broadening of the two-magnon response with doping and temperature has been
studied theoretically in the cuprate
oxides~\cite{Sugai1988,Lyons1988,Lyons1989,Knoll1990,Sulewski1990,Sulewski1991,Blumberg1994,Blumberg1996,Ruebhausen1997,Ruebhausen1999,Sugai2003}.
The two-magnon identifies the strong spin fluctuations in the cuprate even when
the long range antiferromagnetic order disappears at high temperatures or upon
doping~\cite{Devereaux2007}. In a 2D Heisenberg antiferromagnet, the light in
the Raman process flips two nearby spins on the nearest neighboring bond, leaving behind a locally disturbed antiferromagnet
with 6 broken exchange bonds in the final state. Therefore the two-magnon has
the energy scale at roughly $2.7J$~\cite{Devereaux2007}. We equal $2.7J$ to 550~cm$^{-1}$, and obtain the antiferromagnetic exchange strength $J=25$~meV.

\emph{Discussions and Conclusions. --}
The discovery of the superconductivity in the thin film
Nd$_{0.8}$Sr$_{0.2}$NiO$_2$~\cite{Li2019} has caught lots of attention, however,
the recent report of the bulk Nd$_{1-x}$Sr$_{x}$NiO$_2$ ($x=0.2, 0.4$) reveals
an insulating behavior without  the presence of the superconductivity~\cite{Li2019a}. The reason of the
discrepancy is unclear. We suspect that the Sr substitution of Nd  destroys the
flat NiO$_2$ plane in the bulk Nd$_{1-x}$Sr$_{x}$NiO$_2$, probably resulting in the
insulating behavior due to the disorder. In the thin film
Nd$_{1-x}$Sr$_{x}$NiO$_2$, due to the strain effect of substrate, the NiO$_2$
plane remains flat after doping.


In our previous theoretical work, we derive the effective Hamiltonian for nickelate oxides
Nd$_{1-x}$Sr$_x$NiO$_2$ with the flat NiO$_2$ planes~\cite{Zhang2019}. From the the Heyd-Scuseria-Ernzerhof hybrid density functional and the
exact diagonalization of the three-band Hubbard model for the Ni$_5$O$_{16}$
cluster, we
find that the undoped NiO$_2$ plane is a Mott-Hubbard insulator and the doped
holes primarily locate on Ni sites. The Mott-Hubbard scenario is also stressed
in Ref.~\cite{Jiang2019}. The physics of the undoped NiO$_2$ plane
is described by the two-dimensional Heisenberg antiferromagnet on the square
lattice $H=J\sum_{\langle ij \rangle}\mathbf{S}_i\cdot\mathbf{S}_j$ 
with the exchange strength  $J=29$~meV~\cite{Zhang2019}. 

In summary, we perform core-level X-ray photoemission spectroscopy (XPS) and Raman spectroscopy studies of electronic structures and spin
fluctuations in the bulk NdNiO$_2$. We find  that the NiO$_2$ planes in
NdNiO$_2$ are of Mott-Hubbard type in the ZSA scheme, consistent with  Hepting's electronic structure studies on the
thin-film NdNiO$_2$~\cite{Hepting2019}. Two-magnon peak in the Raman scattering
provides direct evidence for the strong spin-fluctuation in NdNiO$_2$. The peak
position determines the antiferromagnetic exchange $J=25$~meV. The present
experimental investigation agrees well with our previous theoretical description
of the electronic structures of NdNiO$_2$.

In the last, we make a remark on  the three-dimensional electron pockets of Nd$^{3+}$ $5d$
character in NdNiO$_2$ as
demonstrated in Ref.~\cite{Lee2004,Hepting2019}. The presence of $5d$ electron
pocket  changes the hole count in the NiO$_2$ planes,  resulting
in spin-$S=0$
Ni$^{2+}$ states in the NiO$_2$ planes even without chemical
doping~\cite{Zhang2019}.  A weakly-interacting three-dimensional 5$d$ metallic state hybridizes with
 the Ni$^+$ local moments in the NiO$_2$ layers, explaining the resonant
 inelastic x-ray scattering spectra in LaNiO$_2$ and NdNiO$_2$~\cite{Hepting2019}.
In our Ni $2p_{3/2}$ XPS spectra, the main line for
NdNiO$_2$ has some weight around 853~eV (Fig.~\ref{fig:figure4}). We can't rule out
the minority nickel metal origin for this weight. However, XRD has the similar
penetration depth as XPS and detect a tiny nickel peak in our samples. So the
853~eV weight in Fig.~\ref{fig:figure4} for NdNiO$_2$ is likely to be intrinsic
and related to the $S=0$ Ni$^{2+}$ states in the NiO$_2$ planes.



\acknowledgements{\textit{Acknowledgments --} 
  J.W.M thanks A. Ng and W.~Q. Chen for useful discussions. J.W.M was partially supported by the program for Guangdong Introducing
  Innovative and Entrepreneurial Teams (No.~2017ZT07C062). This work was also supported by
  National Science Foundation of China (No.~11774143).}

\bibliography{../HTS}
\end{document}